\documentclass
[superscriptaddress,secnumarabic,amssymb,amsmath,nobibnotes,aps,prd,showkeys,showpacs,nofootinbib,onecolumn]{revtex4}%
\usepackage{graphicx}
\usepackage{epstopdf}
\usepackage{bm}
\usepackage{amsmath}
\usepackage{amsfonts}
\usepackage{amssymb}%
\providecommand{\U}[1]{\protect\rule{.1in}{.1in}}
\newcommand{\be}{\begin{equation}}
\newcommand{\ee}{\end{equation}}

\newcommand{\mincir}{\raise
-3.truept\hbox{\rlap{\hbox{$\sim$}}\raise4.truept\hbox{$<$}\ }}
\newcommand{\magcir}{\raise
-3.truept\hbox{\rlap{\hbox{$\sim$}}\raise4.truept\hbox{$>$}\ }}

\begin{document}
\title{Reconstructions of the dark-energy equation of state and the inflationary potential}
\author{John D. Barrow}
\email{J.D.Barrow@damtp.cam.ac.uk}
\affiliation{DAMTP, Centre for Mathematical Sciences, University of Cambridge, Wilberforce
Rd., Cambridge CB3 0WA, UK}
\author{Andronikos Paliathanasis}
\email{anpaliat@phys.uoa.gr}
\affiliation{Instituto de Ciencias F\'{\i}sicas y Matem\'{a}ticas, Universidad Austral de
Chile, Valdivia, Chile}
\affiliation{Department of Mathematics and Natural Sciences, Core Curriculum Program,
Prince Mohammad Bin Fahd University, Al Khobar 31952, Kingdom of Saudi Arabia}
\affiliation{Institute of Systems Science, Durban University of Technology, PO Box 1334,
Durban 4000, Republic of South Africa}

\begin{abstract}
We use a mathematical approach based on the constraints systems in order to
reconstruct the equation of state and the inflationary potential for the
inflaton field from the observed spectral indices for the density
perturbations $n_{s}$ and the tensor to scalar ratio $r$. From the
astronomical data, we can observe that the measured values of these two
indices lie on a two-dimensional surface. We express these indices in terms of
the Hubble slow-roll parameters and we assume that $n_{s}-1=h\left(  r\right)
$. For the function $h\left(  r\right)  $, we consider three cases, where
$h\left(  r\right)  $ is constant, linear and quadratic, respectively. From
this, we derive second-order equations whose solutions provide us with the
explicit forms for the expansion scale-factor, the scalar-field potential, and
the effective equation of state for the scalar field. Finally, we show that
for there exist mappings which transform one cosmological solution to another
and allow new solutions to be generated from existing ones.

\end{abstract}
\date{\today}
\maketitle

\section{Introduction}

An 'inflaton' is a scalar field that can drive a period of acceleration in the
early universe. Such a finite period of inflation \cite{Aref1,guth} can solve
long-standing problems about the structure of the universe that would
otherwise require special initial conditions \cite{planck2013,planck2015}. An
inflaton provides a matter source that can display antigravitating behavior
and so it could also be a candidate for the so-called the 'dark energy' that
drives cosmological acceleration today. It is possible that these two eras of
cosmological acceleration are connected, but so far there is no compelling
theory about how that link might arise between two such widely separated
energy scales.

Various inflationary self-interaction potentials for the inflaton have been
proposed in the literature. Since they lead to different inflationary
scenarios, particularly in respect of the density fluctuations produced, they
have different observational consequences for the cosmic microwave background
radiation, and this permits them to be finely constrained by observational
data. Various inflaton potentials in general relativistic scalar field
cosmology have been proposed in
\cite{ref1a,ref1,ref2,ref3,newinf,ref4,ref5a,ref5,ref6a,ref6,ref7,ref8,ref9,charters}%
, while for inflationary models in other gravity theories, where there are
more possibilities, see
\cite{Aref1,Aref2,Aref3,Aref3b,Aref5,Aref6,Aref6b,Aref7a,Aref7,Aref8,Aref9,Aref10,Aref11}
and references therein.

The construction of the inflaton scalar field potential from observational
data is an open problem of special interest. It provides critical information
about the details of the allowed inflationary models and might provide clues
as to the identity of the inflaton. In
\cite{cp90,cp93,turner1,Adfre,Cos95a,hoi,bpl}, the perturbative reconstruction
approach was applied: the inflaton self-interaction potential, $V\left(
\phi\right)  $,$~$of the scalar field, $\phi$, was reconstructed by
considering a series expansion around a point $\phi=\phi_{0}$, where the
coefficients of the series expansion for the potentials are determined from
the observable values of the scalar spectral index and the usual slow-roll
parameters; for more details see \cite{lidsey}. Alternative approaches to the
reconstruction of the scalar field potential include a stochastic perturbative
approach in \cite{easther}, or another perturbative approach in
\cite{urenalopez}. Two alternative methods for the reconstruction of the
scalar field potential have been proposed in \cite{star} and \cite{Wohns}.
Specifically, in the latter work, an exponential of the scalar field's Hubble
function was considered and found to offer an efficient way to derive and
constrain the power-spectrum observables \cite{Wohns}. By contrast, in
\cite{star}, the scalar field potential was reconstructed for the
Harrison-Zeldovich spectrum by solving the gravitational field equations along
with the equation for the adiabatic scalar perturbations.

The slow-roll parameters and their relations to the spectral indices have been
reconstructed in closed-form \cite{Vallinotto,Chiba,Lin}. This is the approach
that we will follow here to find the equation of state for the effective
perfect fluid which corresponds to the scalar field with a self-interaction
potential.\ While this approach is not so accurate as the previous approaches
(because it depends on approximate relations between the spectral indices and
the slow-roll parameters \cite{lidsey}) it can more easily reconstruct
closed-form solutions for the inflationary potential and the expansion scale
factor expansion. Furthermore, as we shall see in the first approximations for
the models that we study, there exist mappings which transform the models to
other equivalent models and their linearised fluctuations to the
Harrison-Zeldovich spectrum. The plan of this paper is as follows.

In Section \ref{field}, we review scalar field cosmology in a spatially flat
Friedmann-Lema\^{\i}tre-Robertson-Walker (FLRW) universe and introduce the
basic quantities and notations. In Section \ref{section3}, we assume that the
spectral index for the density perturbations, $n_{s}$, and the
tensor-to-scalar ratio, $r$, are related by a function such that
$n_{s}-1=h\left(  r\right)  $. For the defining function, $h\left(  r\right)
,$ we assume that it is either constant, linear or quadratic in $r$. Moreover,
using the slow-roll expressions for these indices, we find ordinary
differential equations whose solutions provide us with the inflationary scalar
field potentials and the equation of state for the energy density and the
pressure of the scalar field while the density perturbations to
tensor-to-scalar ration diagrams are presented for the analytical solutions
that we derive. Moreover, in Section \ref{escape} the values for the free
parameters of the models are determined in order a late time attractor to
exists such that the universe to escape from the inflation phase. Moreover, a
transformation which relates the different models that we study is presented
in Section \ref{section4}. We show that our master equations are all maximally
symmetric. This ensures that maps exist which can transform the solution of
one inflationary model into another. This can be used to determine new
inflationary solutions from known ones. A discussion of the results presented
and our conclusions are given in the concluding Section \ref{conc}.

\section{Underlying equations and definitions}

\label{field}

We take the gravitational field equations to be (with units $8\pi
G=c=\hslash=1$)%
\begin{equation}
G_{\mu\nu}=T_{\mu\nu}^{\left(  \phi\right)  }+T_{\mu\nu}^{\left(  m\right)  },
\label{in.01}%
\end{equation}
where $G_{\mu\nu}=$ $R_{\mu\nu}-\frac{1}{2}g_{\mu\nu}R$ is the Einstein
tensor, $T_{\mu\nu}^{\left(  \phi\right)  }$ is the energy-momentum tensor for
the scalar field,
\begin{equation}
T_{\mu\nu}^{\left(  \phi\right)  }=\phi_{;\mu}\phi_{;\nu}-g_{\mu\nu}\left(
\frac{1}{2}\phi^{;\sigma}\phi_{;\sigma}-V\left(  \phi\right)  \right)  ,
\label{in.01a}%
\end{equation}
and $T_{\mu\nu}^{\left(  m\right)  }$ denotes the energy-momentum tensor of
the other matter sources. Now, we will assume that the universe contains only
the scalar field, so $T_{\mu\nu}^{\left(  m\right)  }=0$. In addition, we have
the propagation equation for the scalar field, $\phi$, from the Bianchi
identity $T_{~~~~~~~\ ;\nu}^{\left(  \phi\right)  \mu\nu}=0,$ which is%
\begin{equation}
-g^{\mu\nu}\phi_{;\mu\nu}+V_{,\phi}=0. \label{in.02}%
\end{equation}

For a spatially-flat FLRW universe, with scale factor, $a\left(  t\right)  $,
the field equations (\ref{in.01})-(\ref{in.02}) are%
\begin{equation}
3H^{2}=\frac{1}{2}\dot{\phi}^{2}+V(\phi), \label{in.03}%
\end{equation}%
\begin{equation}
2\dot{H}+3H^{2}=-\frac{1}{2}\dot{\phi}^{2}+V(\phi), \label{in.04}%
\end{equation}
and
\begin{equation}
\ddot{\phi}+3H\dot{\phi}+V_{,\phi}=0, \label{in.05}%
\end{equation}
where $H=\frac{\dot{a}}{a}$ is the Hubble function and overdots denote
differentials with respect to comoving proper time, $t$. The comoving
observers have $u^{\mu}=\delta_{0}^{\mu}$ , so $u^{\mu}u_{\mu}=-1$. The FLRW
symmetries ensure $\phi=\phi\left(  t\right)  $.

From (\ref{in.01a}), we find that the energy density of the scalar field for
the comoving observer is
\begin{equation}
\rho_{\phi}\equiv\frac{1}{2}\dot{\phi}^{2}+V(\phi); \label{in.06}%
\end{equation}
the pressure is $P_{\phi}=w_{\phi}\rho_{\phi}$, where $w_{\phi}$ is the
equation of state parameter (EoS):
\begin{equation}
w_{\phi}=\frac{\dot{\phi}^{2}-2V(\phi)}{\dot{\phi}^{2}+2V(\phi)}.
\label{in.07}%
\end{equation}

The deceleration parameter, $q$, is given by the formula $q=\frac{1}{2}\left(
1+3w_{\phi}\right)  $ because, as the only matter source is the scalar field,
we have $w_{tot}=w_{\phi}$. The expansion of the universe is accelerated when
$q<0$, that is, $w_{\phi}<-\frac{1}{3}$. Since $V\left(  \phi\right)  >0$, a
negative negative EoS parameter means that the potential dominates the kinetic
term i.e., $\frac{\dot{\phi}^{2}}{2}<V(\phi)$. Furthermore, in the limit
$\dot{\phi}\rightarrow0$ expression (\ref{in.07}) gives $w_{\phi}%
\rightarrow-1$, and the scalar field mimics the cosmological constant.

The so-called potential slow-roll parameters (PSR),
\begin{equation}
\varepsilon_{V}=\left(  \frac{V_{,\phi}}{2V}\right)  ^{2}~\,,~\eta_{V}%
=\frac{V_{,\phi\phi}}{2V}, \label{in.08}%
\end{equation}
have been introduced \cite{slp01} in order to study the existence of the
inflationary phase of the universe. Specifically, the condition for an
inflationary universe is $\varepsilon_{V}<<1$, while in order for the
inflationary phase to last long enough we require the second PSR parameter
also to be small, $\eta_{V}<<1$.

Similarly, the Hubble slow-roll parameters (HSR) have been defined by
\cite{slpv,slp}\qquad%
\begin{equation}
\varepsilon_{H}=-\frac{d\ln H}{d\ln a}=\left(  \frac{H_{,\phi}}{H}\right)
^{2}, \label{in.09}%
\end{equation}
and
\begin{equation}
\eta_{H}=-\frac{d\ln H_{,\phi}}{d\ln a}=\frac{H_{,\phi\phi}}{H}. \label{in.10}%
\end{equation}

It has been shown that the HSR slow-roll parameters are more accurate
descriptors of inflation than the PSR parameters. However, the PSR and HSR
parameters are related and, when $\varepsilon_{H}$ and $\eta_{H}$ are small,
these relations become%
\begin{equation}
\varepsilon_{V}\simeq\varepsilon_{H}~\text{and~}\eta_{V}\simeq\varepsilon
_{H}+\eta_{H}. \label{in.11}%
\end{equation}
In the following we choose to work with the HSR parameters.

\subsection{Analytical solution}

Recently, in ref. \cite{dimakis}, it was found that the field equations,
(\ref{in.03})-(\ref{in.05}), under the transformation, $dt=\exp\left(
\frac{F\left(  \omega\right)  }{2}\right)  d\omega$ with $\omega=6\ln a,$
where $a(t)$ is the cosmic scale factor, can be solved for the scalar field,
$\phi$, and the potential, $V\left(  \phi\right)  ,$ by the following
formulae\footnote{Where a prime \textquotedblleft$^{\prime}$\textquotedblright%
\ denotes the total derivative with respect to $\omega$.}%
\begin{equation}
\phi(\omega)=\pm\frac{\sqrt{6}}{6}\int\!\!\sqrt{F^{\prime}(\omega)}%
d\omega\label{in.12}%
\end{equation}
and
\begin{equation}
V(\omega)=\frac{1}{12}e^{-F(\omega)}\left(  1-F^{\prime}(\omega)\right)  ,
\label{in.13}%
\end{equation}
so the line-element for the FLRW spacetime is now
\begin{equation}
ds^{2}=-e^{F\left(  \omega\right)  }d\omega^{2}+e^{\omega/3}(dx^{2}%
+dy^{2}+dz^{2}). \label{in.14}%
\end{equation}

For the latter line element, the Hubble function is defined as$~H\left(
\omega\right)  =\frac{1}{6}e^{-\frac{F}{2}}$, from where with the use of
(\ref{in.12}) it follows $\frac{dH}{d\phi}=\pm\frac{\sqrt{6}}{2}e^{-\frac
{F}{2}}F^{\prime}\,\,,$ then expression (\ref{in.14}) reduces to the
Hamilton-Jacobi like equation, for $H\left(  \phi\right)  $,
\[
2\left(  \frac{dH\left(  \phi\right)  }{d\phi}\right)  ^{2}-3\left(  H\left(
\phi\right)  \right)  ^{2}+V\left(  \phi\right)  =0,
\]
which studied in \cite{mos1,mos2}.

In the new variables, the effective fluid components for the scalar field are%
\begin{equation}
\rho_{\phi}(\omega)=\frac{1}{12}e^{-F(\omega)}~,~P_{\phi}(\omega)=\frac{1}%
{12}e^{-F(\omega)}\left(  2F^{\prime}(\omega)-1\right)  , \label{in.15}%
\end{equation}
and the effective EoS parameter takes the simple form%
\begin{equation}
w_{\phi}\left(  \omega\right)  =\left(  2F^{\prime}(\omega)-1\right)  .
\label{in.16}%
\end{equation}

These expressions hold for an arbitrary scalar field potential. The field
equations have been reduced to a single first-order differential equation
which can be viewed as a form of the equation of state, $P_{\phi}=P_{\phi
}\left(  \rho_{\phi}\right)  $, for the scalar field. This approach was
applied in \cite{jdband1} in order to construct inflationary potentials from
specific linear and non-linear equations of state.

We can use this solution to express the slow-roll parameters, PSR or HSR, in
terms of the new variable $\omega\equiv\ln(a^{6})$. The HSR parameters are
found to be \cite{jdband1}%
\begin{equation}
\varepsilon_{H}=3F^{\prime}~,~\eta_{H}=3\frac{\left(  F^{\prime}\right)
^{2}-F^{\prime\prime}}{F^{\prime}}, \label{in.17}%
\end{equation}
or, equivalently in terms of the effective EoS parameter,
\begin{equation}
\varepsilon_{H}=\frac{3}{2}\left(  1+w_{\phi}\left(  \omega\right)  \right)  ,
\label{in.18}%
\end{equation}
and
\begin{equation}
\eta_{H}=\frac{3}{2}\frac{\left(  w_{\phi}+1\right)  ^{2}-2w_{\phi,\omega}%
}{\left(  1+w_{\phi}\right)  }. \label{in.19}%
\end{equation}

The number of e-folds is defined to be $N_{e}=\int_{t_{i}}^{t_{f}}H\left(
t\right)  dt=\ln\frac{a_{f}}{a_{i}}=\frac{1}{6}\left(  \omega_{f}-\omega
_{i}\right)  ,$ which means that $N_{e}$ is linearly related to the function
$\omega$. Hence, the slow-roll parameters can be expressed in terms of $N_{e}$.

Lastly, using expression (\ref{in.18}), all the slow-roll parameters can be
expressed in terms of the parameter~$\varepsilon_{H}$ and its derivatives.

\section{Reconstruction of the inflationary potential}

\label{section3}

From the recent data analysis by the Planck 2015 collaboration
\cite{planck2015}, the value of the spectral index for the density
perturbations is $n_{s}=0.968\pm0.006,~$while the range of the scalar spectral
index is $n_{s}^{\prime}=-0.003\pm0.007$. The tensor to scalar ratio, $r$, has
a value smaller than $0.11$, i.e., $r<0.11$.

The mathematical expression which relates the HSR parameters to the spectral
indices $n_{s}$ in the first approximation is%
\begin{equation}
n_{s}\equiv1-4\varepsilon_{H}+2\eta_{H}, \label{in.22}%
\end{equation}
while the tensor to scalar ratio is $r=10\varepsilon_{H}$. Moreover, in the
second approximation the spectral index, $n_{s}$, becomes%
\begin{equation}
n_{s}\equiv1-4\varepsilon_{H}+2\varepsilon_{H}-8\left(  \varepsilon
_{H}\right)  ^{2}\left(  1+2C\right)  +\varepsilon_{H}\eta_{H}\left(
10C+6\right)  -2C\xi_{H}, \label{in.24}%
\end{equation}
where $C=\gamma_{E}+\ln2-2=-0.7296$. So, now it follows that the running index
is%
\begin{equation}
n_{s}^{\prime}\equiv2\varepsilon_{H}\eta_{H}-2\xi_{H}. \label{in.23}%
\end{equation}

From the analysis of the previous section, the spectral indices for the FLRW
spacetime can be written in terms of $\varepsilon_{H}$ and its derivative, or
in terms of the unknown function, $F\left(  \omega\right)  $, and its
derivatives. Recall, that the above expressions for the spectral indices are
definitions and not deductions. However, if we assume that the left-hand side
satisfies some functional expression, i.e., $n_{s}=h\left(  \varepsilon
_{H},...\right)  $,~for an function $h$, then we define a differential
equation, which can be used to construct the exact form for the FLRW spacetime
(\ref{in.14}), i.e., determine $F\left(  \omega\right)  $, that satisfies the
spectral index conditions. Hence, with the use of the solution presented in
the previous section, the scalar field potential can also be derived.

In the following, we consider that
\begin{equation}
n_{s}-1=h\left(  r\right)  , \label{in.23a}%
\end{equation}
and we work with the expression (\ref{in.22}) in the first-order
approximation. Moreover, we assume that we are close to the $n_{s}=1$ spectrum
so that we can treat $h\left(  r\right)  $ as a small correction term to the
spectrum. Hence, the Taylor expansion of the $h\left(  r\right)  $ function
close to a constant value for the scalar ratio, that is, $r=r_{0}$, yields%
\begin{equation}
h\left(  r\right)  =h\left(  r_{0}\right)  +h^{\prime}\left(  r_{0}\right)
\left(  r-r_{0}\right)  +\frac{h^{\prime\prime}\left(  r_{0}\right)  }%
{2!}\left(  r-r_{0}\right)  ^{2}+... \label{in.23b}%
\end{equation}

For our analysis we select three forms for the function $h(r),$ which
\ include the three first terms of the last Taylor expansion for the function
$h\left(  r\right)  $. \ Hence, by substituting from (\ref{in.17}) in
(\ref{in.23a}) three master equations follow for each chosen form of $h\left(
r\right)  $.

\subsection{Constant index: $n_{s}-1=-2n_{0}$}

Assume that the spectral index for the density perturbations is constant, with
$n_{s}-1=-2n_{0}$, where according to the Planck 2015 data at $1\sigma$,
\ $n_{0}$ should be bounded in the range $\,0.013\leq n_{0}\leq0.019$. In the
case where $n_{0}=0$, i.e., $n_{s}=1$, we have the Harrison--Zeldovich
spectrum. These cases were studied before in \cite{Vallinotto,Chiba,Lin}.

\subsubsection{Zero $n_{0}:~$Harrison--Zeldovich spectrum}

Let $n_{0}=0$, so $n_{s}=1~$and we have the exact Harrison-Zeldovich spectrum.
Then, from (\ref{in.22}), it follows that $\eta_{H}=2\varepsilon_{H}$. Hence,
from (\ref{in.17}) the second-order differential equation for $F(\omega)$ is
\begin{equation}
F^{\prime\prime}+\left(  F^{\prime}\right)  ^{2}=0, \label{in.25}%
\end{equation}
which has the solution $F\left(  \omega\right)  =\ln\left(  F_{1}\left(
\omega-\omega_{0}\right)  \right)  $, where the effective equation-of-state
parameter is now
\begin{equation}
w_{\phi}\left(  \omega\right)  =-1+\frac{2}{\omega-\omega_{0}}. \label{in.26}%
\end{equation}

The differential equation, (\ref{in.25}), was derived in \cite{jdband1} and it
follows from the generalized Chaplygin gas \cite{jdbchg} with $\lambda=2$,
that is for an EoS
\begin{equation}
p_{\phi}=\gamma\rho_{\phi}^{\lambda}-\rho_{\phi}~\text{with}~\lambda=2,
\label{in.27}%
\end{equation}
where $\gamma\varpropto F_{1}$. \ Therefore, with the use of expressions
(\ref{in.12})-(\ref{in.14}) we find that in the proper time where $N\left(
t\right)  =1$, \ the scale factor is that of an intermediate inflation
(\cite{star,inter,inter2,BLM}),
\begin{equation}
a\left(  t\right)  \simeq\exp\left(  a_{1}t^{2/3}\right)  , \label{in.28a}%
\end{equation}
and the scalar-field potential is%
\begin{equation}
V\left(  \phi\right)  =\frac{1}{18F_{1}}\left(  \phi^{-2}-\frac{2}{3}\phi
^{-4}\right)  . \label{in.29}%
\end{equation}

Here it is important to mention that the scalar field description of the
inflaton is valid only for values of $\phi$, such that the scalar field
potential is not negative. Note that the non essential integration constants
have been absorb and $\phi$ indicates $\phi-\phi_{0}$, where in (\ref{in.29})
without loss of generality we considered $\phi_{0}=0$.

\subsubsection{Non-zero $n_{0}$}

We now assume that $n_{s}-1=-2n_{0}\neq0$. Then, from (\ref{in.22}), it
follows that $\eta_{H}=2\varepsilon_{H}-n_{0}$ and with the use of
(\ref{in.17}), the differential equation for $F(\omega)$ is now
\begin{equation}
F^{\prime\prime}+\left(  F^{\prime}\right)  ^{2}-\frac{n_{0}}{3}F^{\prime}=0,
\label{in.30}%
\end{equation}
with the closed-form solution%
\begin{equation}
F\left(  \omega\right)  =\ln\left\{  F_{1}\exp\left(  \frac{n_{0}}{3}%
\omega\right)  \right\}  +F_{0}. \label{in.31}%
\end{equation}

The latter function has been derived in \cite{jdbchg} as the solution in which
the scalar field mimics the generalized Chaplygin gas (or a bulk viscosity)
with EoS parameter
\begin{equation}
p_{\phi}=A\rho_{\phi}^{2}+B\rho_{\phi} \label{in.32}%
\end{equation}
\qquad for the specific values $A,B$ such that $F_{0}=-\frac{A}{1+B}$ and
$B=1+\frac{2}{3}n_{0}$.

Furthermore, the effective EoS parameter is calculated to be%
\begin{equation}
w_{\phi}\left(  \omega\right)  =-1+\frac{n_{0}}{3}-\frac{F_{0}n_{0}}{3}\left(
F_{1}\exp\left(  \frac{n_{0}}{3}\omega\right)  +F_{0}\right)  ^{-1},
\label{in.33}%
\end{equation}
while the closed-form expression for the scalar field potential is%
\begin{equation}
V\left(  \phi\right)  =\frac{F_{1}}{9}\frac{e^{\sqrt{\frac{n_{0}}{3}}\phi}%
}{\left(  e^{\sqrt{\frac{n_{0}}{3}}\phi}+F_{0}F_{1}\right)  ^{2}}\left(
3-n_{0}\left(  \frac{e^{\sqrt{\frac{n_{0}}{3}}\phi}-F_{0}F_{1}}{e^{\sqrt
{\frac{n_{0}}{3}}\phi}+F_{0}F_{1}}\right)  ^{2}\right)  . \label{in.34}%
\end{equation}
The expansion scale factor cannot be written in a closed-form expression in
the proper time, $t$. Moreover, for the potential (\ref{in.34}), we have that
for large values of $\phi$, the potential becomes exponential, that is,%

\begin{equation}
\lim_{\phi\rightarrow+\infty}V\left(  \phi\right)  =\frac{\left(
3-n_{0}\right)  F_{1}}{9}e^{-\sqrt{\frac{n_{0}}{3}}\phi},
\end{equation}
and approximates the solution in which the scalar field mimics a perfect fluid
with constant equation of state parameter. In order to determine the physical
properties of the parameter $n_{0}$, but also those of the integration
constants $F_{0}$ and $F_{1}$, the indices $n_{s}$ and $r$ are calculated below.

\subsubsection{Observational constraints}

For the solution (\ref{in.31}), we calculate the slow-roll parameters to be
\begin{equation}
\varepsilon_{H}=n_{0}\left(  1-\left(  1+\frac{F_{1}}{F_{0}}e^{\frac{n_{0}}%
{3}\omega}\right)  ^{-1}\right)  ~,~\eta_{H}=2\varepsilon_{H}-n_{0},
\label{oc.01}%
\end{equation}
which gives that $n_{s}-1=-2n_{0}.~$Recall that inflation ends at $\omega
_{f},$ where $\varepsilon_{H}\left(  \omega_{f}\right)  =1$. Hence, we find
that
\begin{equation}
\omega_{f}=\frac{3}{n_{0}}\ln\left[  \frac{F_{0}}{F_{1}\left(  n_{0}-1\right)
}\right]  , \label{oc.02}%
\end{equation}
and
\begin{equation}
n_{s}\left(  n_{0},N_{e}\right)  -1=-2n_{0}~,~r\left(  n_{0},N_{e}\right)
=\frac{10n_{0}}{1+\left(  n_{0}-1\right)  e^{-2n_{0}N_{e}}}, \label{oc.03}%
\end{equation}
where $N_{e}$ is the number of e-folds; recall that $6N_{e}=\omega_{f}%
-\omega_{i}.$

\begin{figure}[ptb]
\includegraphics[height=7cm]{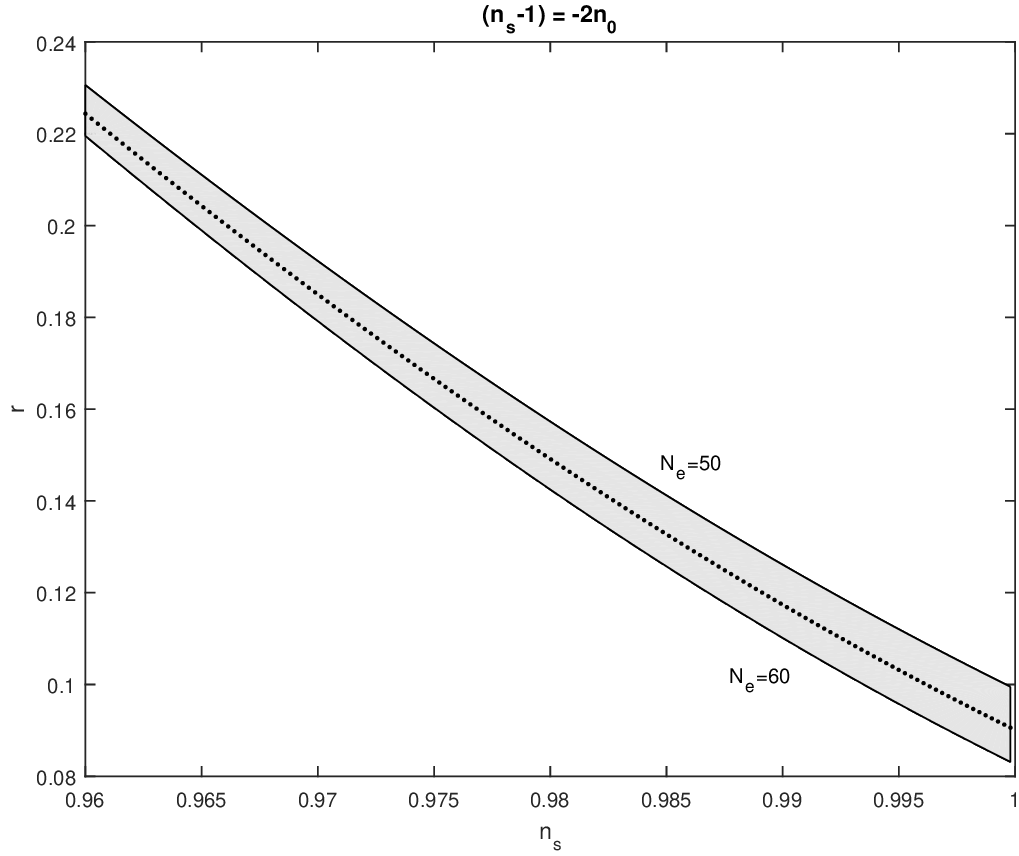}
%\mbox{\epsfxsize=14.2cm \epsffile{dustplusminus.eps}}
\caption{Spectral index $n_{s}$ to scalar to tensor ratio $r\,,$ for the
scalar field potential in which $n_{s}-1=-2n_{0}.~$The figure is for various
values of $n_{0}$ in the range $n_{0}\in\left(  0,0.02\right)  $ and for
number of e-folds $N_{e}\in\left[  50,60\right]  $. The dot line is for
$N_{e}=55$. }%
\label{linear01}%
\end{figure}

From the latter expressions, it follows that while the value of $n_{0}$ fixes
the index $n_{s}$, only the scalar-tensor ratio $r$ depends on the number of
e-folds. Furthermore, the integration constants are non-essential and fix the
value of the scale factor at the end of the inflation. In Figure
\ref{linear01} the $n_{s}-r$ plot is presented for the expressions
(\ref{oc.03}) and for $n_{0}\in\left(  0,0.02\right)  $,~$\ N_{e}\in\left[
50,60\right]  $. Note that for values of $n_{0}$ where $n_{s}$ is constrained
by the Planck 2015 data, it follows that $r<0.11$ for very large values of
$N_{e}$, \ while for the number of e-folds that we considered in the figure
$r>0.11$.

Furthermore, in the case of the Harrison-Zeldovich spectrum, that is,
$n_{0}=0$, we calculate $r=\frac{10}{1-2N_{e}}$, hence $r<0.11$ when
$N_{e}>50$. As before, the integration constants (now $F_{1}$ and $\omega_{0}%
$) specify only the value of $\omega_{f}$ at which inflation ends. We
calculate that $\omega_{f}=3+\omega_{0}$.

\subsection{Linear expression: $n_{s}-1\simeq r$}

We continue now by taking the more general ansatz, $n_{s}-1=-2n_{1}%
\varepsilon_{H}-2n_{0}$; that is, the spectral index $n_{s}$ depends linearly
on the tensor to scalar ratio,~$r$. Recall that $r=10\varepsilon_{H}$; so in
the limit in which $n_{1}\rightarrow0$ we are in a situation where
$n_{s}-1=~const$. \ We study two cases: $n_{0}=0$ and $n_{0}\neq0$.

\subsubsection{Zero $n_{0}:$}

When $n_{0}=0$, so $n_{s}=1$ and $r=0$, with the use of (\ref{in.17}) the
differential equation for $F(\omega)$ is%
\begin{equation}
F^{\prime\prime}+\left(  1-n_{1}\right)  \left(  F^{\prime}\right)  ^{2}=0,
\label{in.35}%
\end{equation}
with solution%
\begin{equation}
F\left(  \omega\right)  =-\frac{1}{n_{1}-1}\ln\left(  F_{1}\left(
\omega-\omega_{0}\right)  \right)  ~,~n_{1}\neq1, \label{in.36}%
\end{equation}
or
\begin{equation}
F\left(  \omega\right)  =F_{1}\left(  \omega-\omega_{0}\right)  ,~n_{1}=1,
\label{in.37}%
\end{equation}
where $F_{1}$ is constant. Of course, one should be careful because we have
assumed that $n_{s}$ is given in terms of the first approximation, i.e.,
$\left(  \varepsilon_{H}\right)  ^{2}\simeq0,$ and second-order approximations
may need to be considered. For simplicity, we continue with the first-order
approximations. The ansatz is stronger if $n_{1}$ is of order $O\left(
\varepsilon_{H}\right)  ^{-1}$.

Obviously, for $n_{1}=0$, eqn. (\ref{in.25}) is recovered. Eqn. (\ref{in.36})
corresponds to the solution of the generalized Chaplygin gas, (\ref{in.27}),
with $\lambda=2-n_{1}.~$ The scalar-field potential is given by the expression
\cite{jdbchg}
\begin{equation}
V\left(  \phi\right)  \varpropto\phi^{-\frac{2}{1-\lambda}}\left(  1-\frac
{2}{3\left(  1-\lambda^{2}\right)  }\phi^{-2}\right)  \label{in.38}%
\end{equation}
and the scale factor is that of intermediate inflation $a\left(  t\right)
\simeq\exp\left(  a_{1}t^{N}\right)  ~$for$~n\neq\frac{3}{2}~$and $a\left(
t\right)  \simeq\exp\left(  a_{1}e^{\bar{\gamma}t}\right)  ~~$for $~n=\frac
{3}{2}$. Moreover, the effective EoS is derived to be
\begin{equation}
w_{\phi}\left(  \omega\right)  =-1+\frac{2}{\left(  1-n_{1}\right)  }\left(
\omega-\omega_{0}\right)  ^{-1}. \label{in.39}%
\end{equation}

In the limit where $n_{1}=1,$ from solution (\ref{in.37}) we calculate
$w_{\phi}\left(  \omega\right)  =-1+2F_{1}$, which is a particular solution of
the exponential potential $V\left(  \phi\right)  =\frac{\left(  1-F_{1}%
\right)  }{12}e^{-\sqrt{6F_{1}}\phi}$, and there is a power-law scale
factor~$a\left(  t\right)  \varpropto t^{\frac{1}{3F_{1}}}$.

\subsubsection{Non-zero $n_{0}:$}

Now we assume that $n_{0}\neq0$. The unknown function, $F\left(
\omega\right)  $, which provides the solution for the spacetime metric,
satisfies the second-order nonlinear differential equation%
\begin{equation}
F^{\prime\prime}+\left(  1-n_{1}\right)  \left(  F^{\prime}\right)  ^{2}%
-\frac{n_{0}}{3}F^{\prime}=0 \label{in.40}%
\end{equation}
with closed-form solution%
\begin{equation}
F\left(  \omega\right)  =-\frac{1}{n_{1}-1}\ln\left(  F_{1}\exp\left(
\frac{n_{0}}{3}\omega\right)  +F_{0}\right)  ,~n_{1}\neq1, \label{in.41}%
\end{equation}
or%
\begin{equation}
F\left(  \omega\right)  =F_{1}\exp\left(  \frac{n_{0}}{3}\omega\right)
+F_{0}~,~n_{1}=1. \label{in.42}%
\end{equation}

As in the case of $n_{0}=0$, when $n_{1}\neq1$ the solution generalizes that
of (\ref{in.31}) and the scalar field now satisfies the equation of state of
the generalized Chaplygin gas, namely
\begin{equation}
p_{\phi}=A\rho_{\phi}^{\lambda}+B\rho_{\phi}, \label{in.42a}%
\end{equation}
where, in contrast to (\ref{in.32}) where $\lambda=2$, we now have
$\lambda=2-n_{1}$.

Again, the scalar-field potential is given in terms of the hyperbolic
functions as \cite{jdband1}
\begin{equation}
V\left(  \phi\right)  =\frac{1}{36}\left(  \frac{e^{-\sqrt{\frac{n_{0}\left(
\lambda-1\right)  }{3}}\phi}}{4F_{1}}\left(  1+e^{\sqrt{\frac{n_{0}\left(
\lambda-1\right)  }{3}}\phi}\right)  ^{2}\right)  ^{\frac{1}{1-\lambda}%
}\left(  3-\frac{n_{0}}{\lambda-1}\left(  \frac{e^{\sqrt{\frac{n_{0}\left(
\lambda-1\right)  }{3}}\phi}-F_{0}F_{1}}{e^{\sqrt{\frac{n_{0}\left(
\lambda-1\right)  }{3}}\phi}+F_{0}F_{1}}\right)  ^{2}\right)  . \label{in.43}%
\end{equation}
Alternatively, for $n_{1}=1,$ the scalar-field potential is
\begin{equation}
V\left(  \phi\right)  =\frac{1}{72}\exp\left(  -\frac{n_{0}}{2}\phi^{2}%
-F_{0}\right)  \left(  6-\left(  n_{0}\phi\right)  ^{2}\right)  .
\label{in.44}%
\end{equation}

The scale factor, $a\left(  t\right)  ,$ cannot be written as a closed-form
expression in either case. However, for the effective EoS parameter we have
\begin{equation}
w_{\phi}\left(  \omega\right)  =-1+\frac{2}{3}n_{0}F_{1}\exp\left(
\frac{n_{0}}{3}\omega\right)  ~,~n_{1}=1 \label{in.45}%
\end{equation}
and%
\begin{equation}
w_{\phi}=-1+\frac{2}{3}\frac{n_{0}F_{1}\exp\left(  \frac{n_{0}}{3}%
\omega\right)  }{1-n_{1}}\left(  F_{1}\exp\left(  \frac{n_{0}}{3}%
\omega\right)  +F_{0}\right)  ^{-1}~,~n_{1}\neq1. \label{in.46}%
\end{equation}

So far, the generalized Chaplygin gas which leads to intermediate inflation,
and another generalization of the Chaplygin gas which was studied in
\cite{jdband1}, have been recovered. For these two inflationary models the
scalar-field potentials have similar forms. For one model the potential,
$V\left(  \phi\right)  $, is given by a polynomial function of $\phi$, while
for the second model it is given as a function of the hyperbolic trigonometric functions.

\subsubsection{Observational constraints}

\begin{figure}[ptb]
\includegraphics[height=7cm]{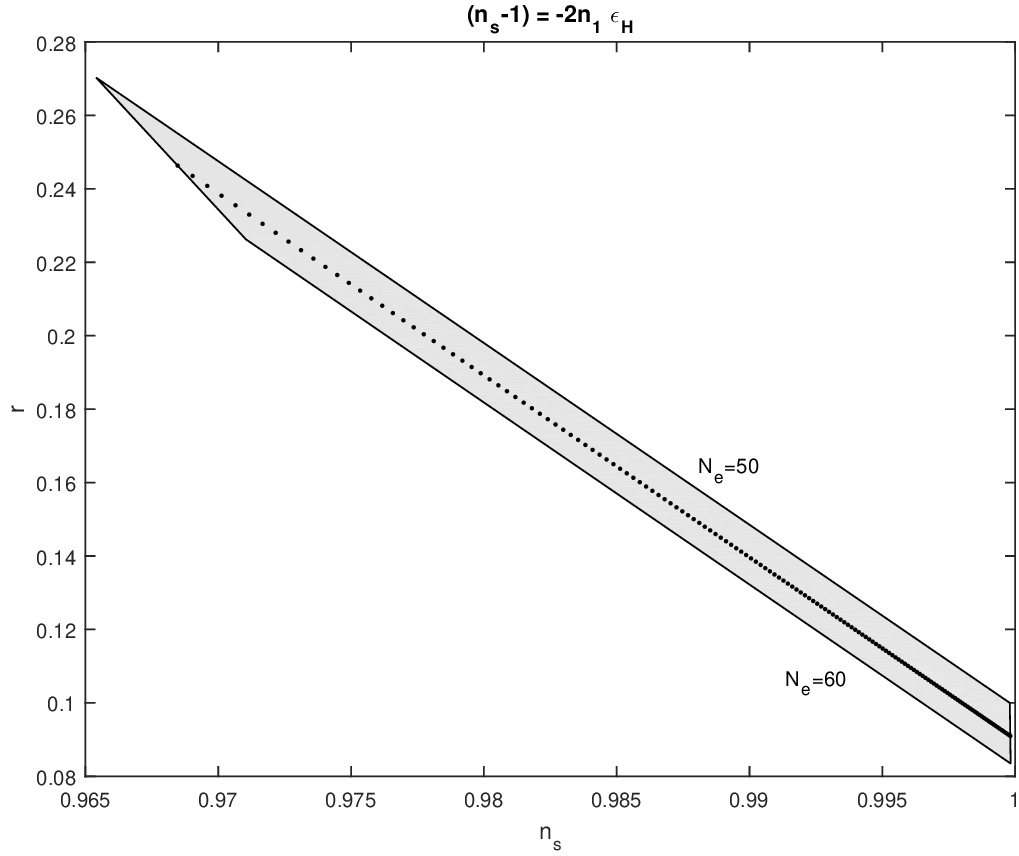}
%\mbox{\epsfxsize=14.2cm \epsffile{dustplusminus.eps}}
\caption{Spectral index $n_{s}$ to scalar to tensor ratio $r\,,$ for the
scalar field potential in which $n_{s}-1=-2n_{1}\varepsilon_{H}.~$The figure
is for various values of $n_{1}$ in the range $n_{1}\in\left(
0.01,0.65\right)  $ and for number of e-folds $N_{e}\in\left[  50,60\right]
$. The dot line is for $N_{e}=55$. }%
\label{fig02}%
\end{figure}

The slow-roll parameters for the solution (\ref{in.36}) are calculated to be%
\begin{equation}
\varepsilon_{H}=\frac{3}{1-n_{1}}\left(  \omega-\omega_{0}\right)
^{-1}~,~\eta_{H}=\left(  n_{1}-2\right)  \varepsilon_{H}. \label{oc.04}%
\end{equation}
Hence, $\omega_{f}=\frac{3}{1-n_{1}}+\omega_{0},$ from which we find%
\begin{equation}
n_{s}\left(  n_{1},N_{e}\right)  =1+\frac{2n_{1}}{2\left(  n_{1}-1\right)
N_{e}-1}~,~r\left(  n_{1},N_{e}\right)  =\frac{10}{1-2\left(  n_{1}-1\right)
N_{e}}. \label{oc.05}%
\end{equation}
As before, the constants of integration have no effect on the inflationary
parameters. \ Furthermore, we see that we have $n_{s}\left(  N_{e}\right)
-1<0$, so necessarily $n_{1}>0$ and $n_{1}<1+\frac{1}{2N_{e}}$, while the
latter also ensures that $r\left(  N_{e}\right)  >0.$ The case in which
$n_{1}=1$ corresponds to the exponential potential and and gives constant
slow-roll parameters. The $n_{s}-r$ diagram for the expressions (\ref{oc.05})
is given in Figure \ref{fig02}, for $N_{e}\in\left[  50,60\right]  $ and the
free parameter $n_{1},$ with $n_{1}\in\left(  0.01,0.65\right)  $.

For $n_{0}\neq0$, from (\ref{in.41}) it follows that%
\begin{equation}
\varepsilon_{H}=\frac{n_{0}}{n_{1}-1}\left(  1+\frac{F_{0}}{F_{1}}%
e^{-\frac{n_{0}}{3}\omega}\right)  ~^{-1}~,~\ \eta_{H}=\frac{n_{0}}{n_{1}%
-1}\left(  \frac{F_{1}e^{\frac{n_{0}}{3}\omega}+F_{0}\left(  n_{1}-1\right)
}{F_{1}e^{\frac{n_{0}}{3}\omega}+F_{0}}\right)  , \label{oc.06}%
\end{equation}
which shows that inflation ends when
\begin{equation}
\omega_{f}=\frac{3}{n_{0}}\ln\left(  \frac{F_{0}}{F_{1}}\frac{\left(
1-n_{1}\right)  }{n_{0}-\left(  1-n_{1}\right)  }\right)  , \label{oc.07}%
\end{equation}
from which we find%
\begin{equation}
n_{s}\left(  n_{0},n_{1},N_{e}\right)  =1-\frac{2n_{0}\left(  1+\left(
n_{0}+n_{1}-1\right)  e^{-2n_{0}N_{e}}\right)  }{\left(  1-n_{1}\right)
+\left(  n_{0}+n_{1}-1\right)  e^{-2n_{0}N_{e}}}~, \label{oc.08}%
\end{equation}%
\begin{equation}
~~r\left(  n_{0},n_{1},N_{e}\right)  =\frac{10n_{0}}{\left(  1-n_{1}\right)
+\left(  n_{0}+n_{1}-1\right)  e^{-2n_{0}N_{e}}}. \label{oc.09}%
\end{equation}

\begin{figure}[ptb]
\includegraphics[height=7cm]{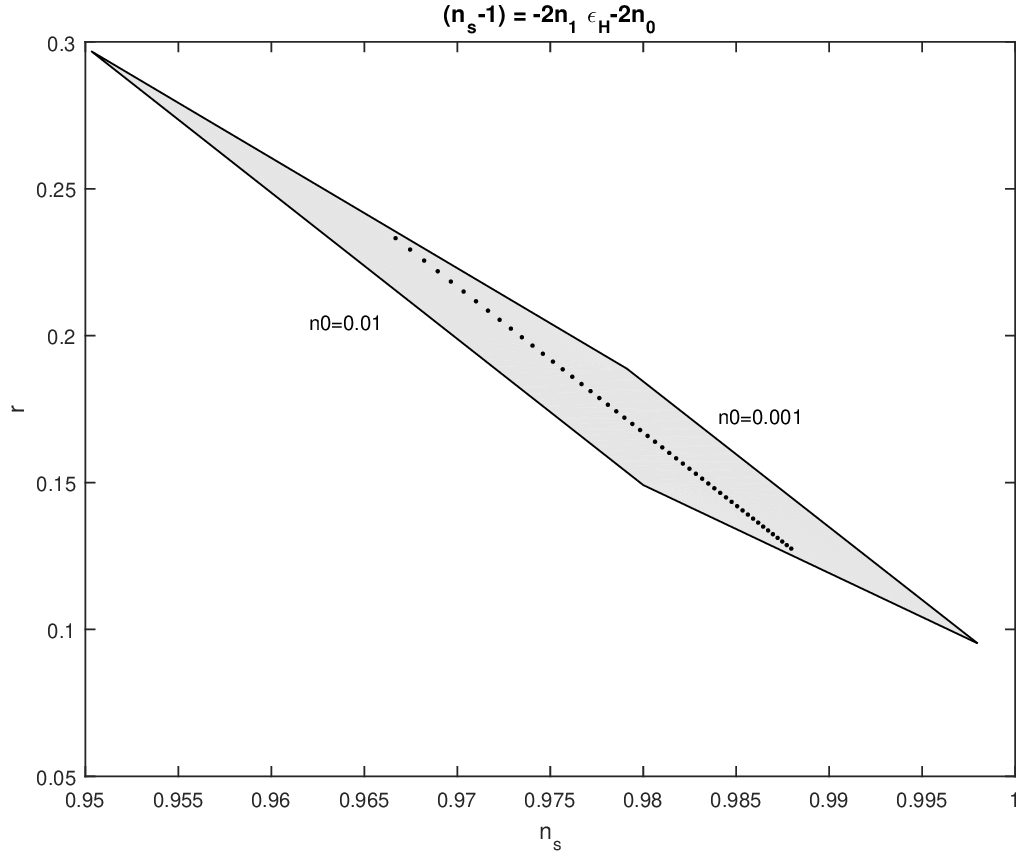}
%\mbox{\epsfxsize=14.2cm \epsffile{dustplusminus.eps}}
\caption{Spectral index $n_{s}$ to scalar to tensor ratio $r\,,$ for the
scalar field potential in which $n_{s}-1=-2n_{1}\varepsilon_{H}-2n_{0}.~$The
figure is for various values of \ the parameters $n_{0}$ and $n_{1};$ in the
range $n_{0}\in\left[  0.001,0.01\right]  $, $n_{1}\in\left[
0.001,0.5\right]  $ and for number of e-folds $N_{e}=55$. The dot line is for
$n_{0}=0.005$. }%
\label{fig03}%
\end{figure}

\begin{figure}[ptb]
\includegraphics[height=7cm]{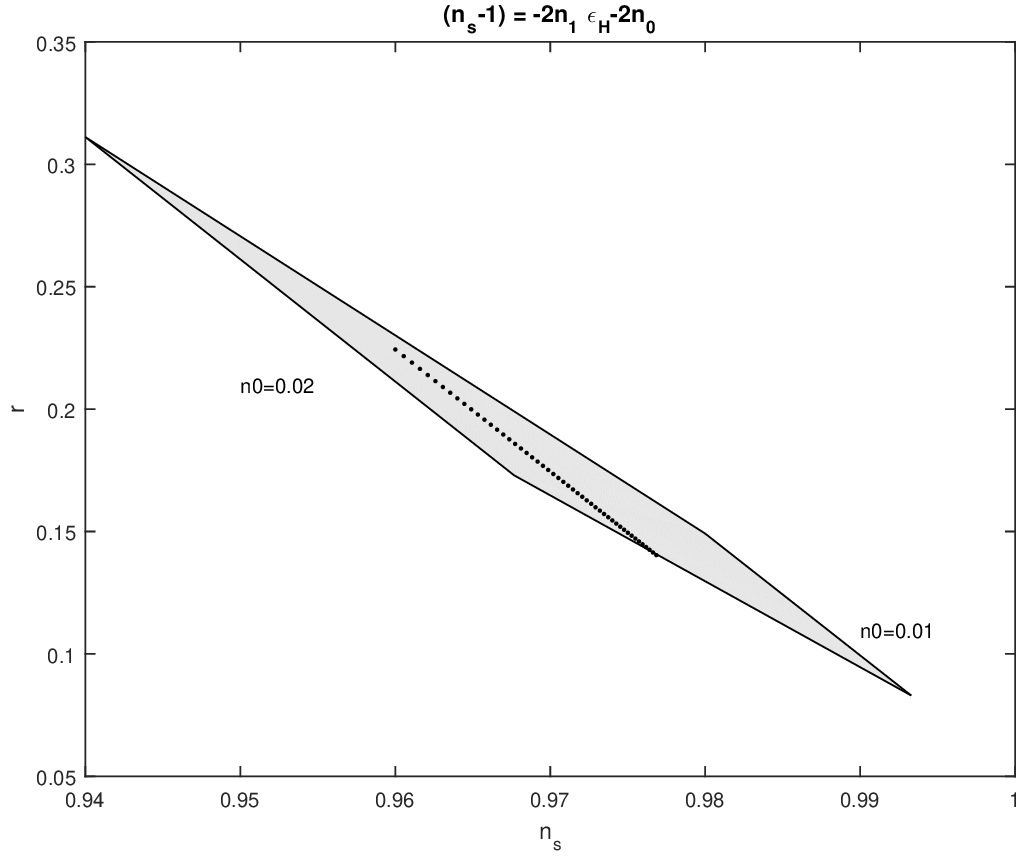}
%\mbox{\epsfxsize=14.2cm \epsffile{dustplusminus.eps}}
\caption{Spectral index $n_{s}$ to scalar to tensor ratio $r\,,$ for the
scalar field potential in which $n_{s}-1=-2n_{1}\varepsilon_{H}-2n_{0}.~$The
figure is for various values of \ the parameters $n_{0}$ and $n_{1};$ in the
range $n_{0}\in\left[  0.01,0.03\right]  $, $n_{1}\in\left[
-0.5,0.001\right]  $ and for number of e-folds $N_{e}=55$. \ The dot line is
for $n_{0}=0.02$.}%
\label{fig04}%
\end{figure}

The $n_{s}-r$ diagram for the parameters (\ref{oc.08}), (\ref{oc.09}) is given
in Figures \ref{fig03} and \ref{fig04}, for the number of e-folds $N_{e}=55$
and for various values of the free parameters $n_{0}$ and $n_{1}$. Figure
\ref{fig03} is for $n_{0}\in\left[  0.001,0.01\right]  $ and $n_{1}\in\left[
0.001,0.5\right]  $, while Figure \ref{fig04} is for $n_{0}\in\left[
0.01,0.03\right]  $ and $n_{1}\in\left[  -0.5,-0.001\right]  $.

We continue our analysis with a more general case in which the relation
between $n_{s}$ and $r$ is parabolic.

\subsection{Parabolic: $n_{s}-1\simeq r^{2}$}

Consider now the case where the relation $n_{s}-1=h\left(  \varepsilon
_{H}\right)  $ describes a parabola such that
\begin{equation}
n_{s}-1=2n_{2}\left(  \varepsilon_{H}\right)  ^{2}-2n_{1}\varepsilon
_{H}-2n_{0}. \label{in.47}%
\end{equation}
From the constraint equation above, the nonlinear differential equation for
$F\left(  \omega\right)  $ is%
\begin{equation}
F^{\prime\prime}+3n_{2}\left(  F^{\prime}\right)  ^{3}+\left(  1-n_{1}\right)
\left(  F^{\prime}\right)  ^{2}-\frac{n_{0}}{3}F^{\prime}=0, \label{in.48}%
\end{equation}
which can be written as a first-order ordinary differential equation in terms
of the effective EoS parameter or in terms of the HSR parameter,
$\varepsilon_{H}$, as
\begin{equation}
3\varepsilon_{H}^{\prime}+n_{2}\left(  \varepsilon_{H}\right)  ^{3}+\left(
1-n_{1}\right)  \left(  \varepsilon_{H}\right)  ^{2}-n_{0}\varepsilon_{H}=0.
\label{in.49}%
\end{equation}
As in the linear case, for completeness, one has to consider the second-order
approximation in the definition of $n_{s}\left(  \varepsilon_{H},\eta
_{H}\right)  $. However, we continue just with the first approximation here.
The ansatz is consistent if $n_{2}~$is of the order $n_{2}\simeq\left(
\varepsilon_{H}\right)  ^{-2}.$

The general solution of eqn. (\ref{in.49}) is%
\begin{equation}
\frac{3}{2n_{0}}\ln\left(  n_{2}+\frac{\left(  1-n_{1}\right)  }%
{\varepsilon_{H}}-\frac{n_{0}}{\left(  \varepsilon_{H}\right)  ^{2}}\right)
-\frac{3\left(  1-n_{1}\right)  \arctan\left(  \frac{2n_{2}\varepsilon
_{H}+\left(  1-n_{1}\right)  }{\sqrt{4n_{0}n_{2}+\left(  1-n_{1}\right)  ^{2}%
}}\right)  }{n_{0}\sqrt{4n_{0}n_{2}+\left(  1-n_{1}\right)  ^{2}}}=\left(
\omega-\omega_{0}\right)  , \label{in.50}%
\end{equation}
which for some specific values of the free parameters can be written in closed form.

In the special case of $n_{1}=1$ we find that
\begin{equation}
\left(  \varepsilon_{H}\right)  ^{2}=\frac{n_{0}}{c_{1}e^{-\frac{2}{3}%
n_{0}\omega}+n_{2}}~,~~n_{0}\neq0, \label{in.51}%
\end{equation}
and
\begin{equation}
\left(  \varepsilon_{H}\right)  ^{2}=\frac{3}{2n_{2}\omega+c_{1}}~,~n_{0}=0.
\label{in.52}%
\end{equation}

Hence, for the function $F\left(  \omega\right)  $ defining the metric, we
have%
\begin{equation}
F\left(  \omega\right)  =\pm\frac{\text{\textrm{arctanh}}\left(  \sqrt
{1+\frac{c_{1}}{n_{2}}}e^{-\frac{2}{3}n_{0}\omega}\right)  }{\sqrt{n_{0}n_{2}%
}}~,~~n_{0}\neq0, \label{in.53}%
\end{equation}
and%
\begin{equation}
F\left(  \omega\right)  =\pm\sqrt{3}\sqrt{\frac{2}{n_{2}}\omega+c_{1}}%
~,~n_{0}=0. \label{in.54}%
\end{equation}

Therefore, from (\ref{in.53}), the potential is found to be%
\begin{equation}
V\left(  \phi\right)  \varpropto\frac{1}{12}\exp\left(  -V_{1}\phi
^{2/3}\right)  \left(  1+V_{2}\phi^{-\frac{2}{3}}\right)  , \label{in.55}%
\end{equation}
where $V_{1,2}=V_{1,2}\left(  n_{2}\right)  $ are constant,$~$and it can be
seen that it has the form of the potential in (\ref{in.44}). For small values
of $\left\vert \phi\right\vert ,$the potential, (\ref{in.55}) becomes the
power-law potential $V\left(  \phi\right)  \simeq\phi^{-\frac{2}{3}}$, which
means that finite-time singularities of the 'generalized sudden' type can
follow \cite{graham}. Moreover, for the EoS for the scalar field it follows
that the effective equation of state is
\begin{equation}
p_{\phi}=\left(  \frac{6\rho_{\phi}}{n_{2}\ln\left(  12\rho_{\phi}\right)
}-\rho_{\phi}\right)  ~,~n_{0}=0. \label{in.56}%
\end{equation}

On the other hand, for $n_{0}\neq0$ we find that that the EoS is
\begin{equation}
p_{\phi}=2\sqrt{\frac{n_{0}}{n_{2}}}\left(  \frac{16n_{0}n_{2}\rho_{\phi}%
^{2}+1}{16n_{0}n_{2}\rho_{\phi}^{2}-1}\right)  -\rho_{\phi}~,~n_{0}\neq0.
\label{in.57}%
\end{equation}
The scalar-field potential for $n_{0}\neq0$ cannot be written in closed form.
However, in terms of $\omega$ it is
\begin{equation}
V\left(  \omega\right)  =e^{-F\left(  \omega\right)  }\left(  1\pm\sqrt
{\frac{n_{0}}{n_{2}}\left(  1+\frac{c_{1}}{n_{2}}e^{-\frac{2}{3}n_{0}\omega
}\right)  ^{-1}}\right)  . \label{in.58}%
\end{equation}

\subsubsection{Observational constraints}

For the solution (\ref{in.54}), in which $n_{0}=0$ and $n_{1}=1$, we find that
inflation ends when $\omega_{f}=\frac{27-c_{1}\left(  n_{2}\right)  ^{2}%
}{2n_{2}}$, and the parameters $n_{s}$ and $r$ are given in terms of the
number of e-folds by
\begin{equation}
n_{s}-1=\frac{2n_{2}-6\sqrt{9+4n_{2}N_{e}}}{9+4n_{2}N_{e}}~,~r=\frac{30}%
{\sqrt{9+4n_{2}N_{e}}}. \label{oc.10}%
\end{equation}

\begin{figure}[ptb]
\includegraphics[height=7cm]{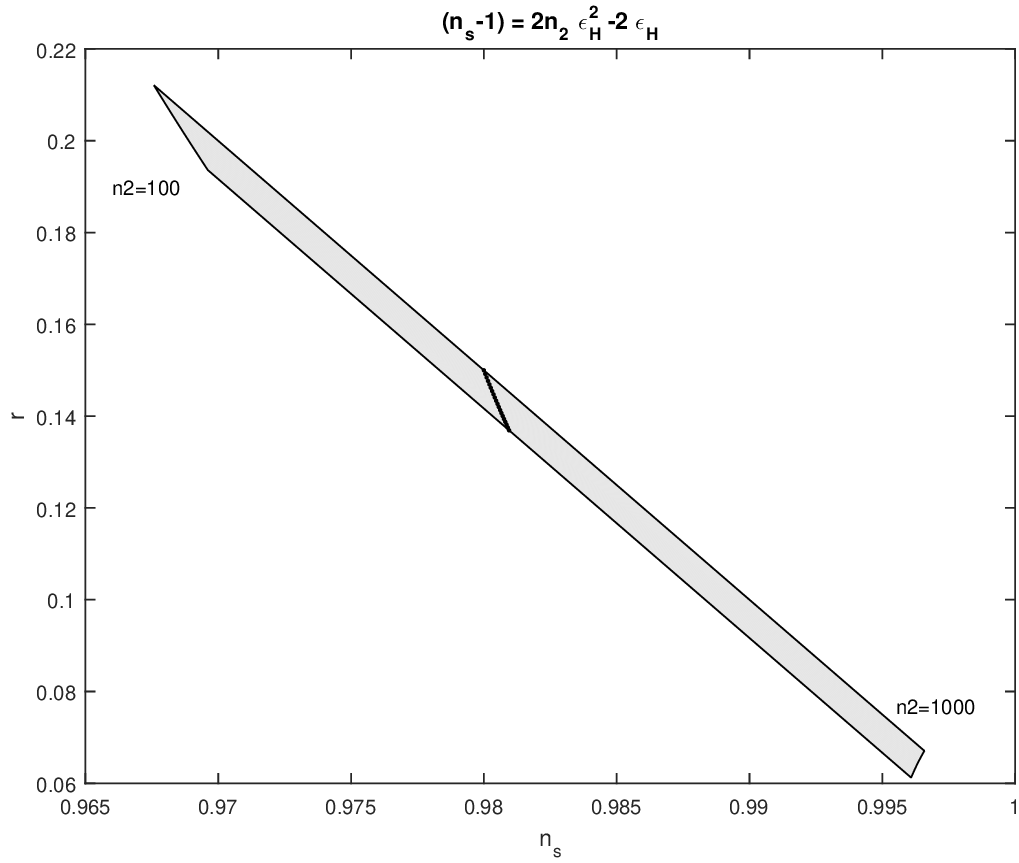}
%\mbox{\epsfxsize=14.2cm \epsffile{dustplusminus.eps}}
\caption{Spectral index $n_{s}$ to scalar to tensor ratio $r\,,$ for the
scalar field potential in which $n_{s}-1=2n_{2}\left(  \varepsilon_{H}\right)
^{2}-2\varepsilon_{H}.~$The figure is for various values of \ the parameter
$n_{2}\in\left(  10^{2},10^{3}\right)  $ and number of e-folds $N_{e}%
\in\left[  50,60\right]  $. The dot line is for $n_{2}=2\times10^{2}$. }%
\label{fig05}%
\end{figure}

In Fig. \ref{fig05} the $n_{s}-r$ diagram is given for the parameters
(\ref{oc.10}) with $n_{2}\in\left(  10^{2},10^{3}\right)  $ and \ $N_{e}%
\in\left[  50,60\right]  $. Note that in order for this case to differ from
the linear we have assumed that $n_{2}$ has a large value of order $\left(
\varepsilon_{H}\right)  ^{-1}$.

\begin{figure}[ptb]
\includegraphics[height=7cm]{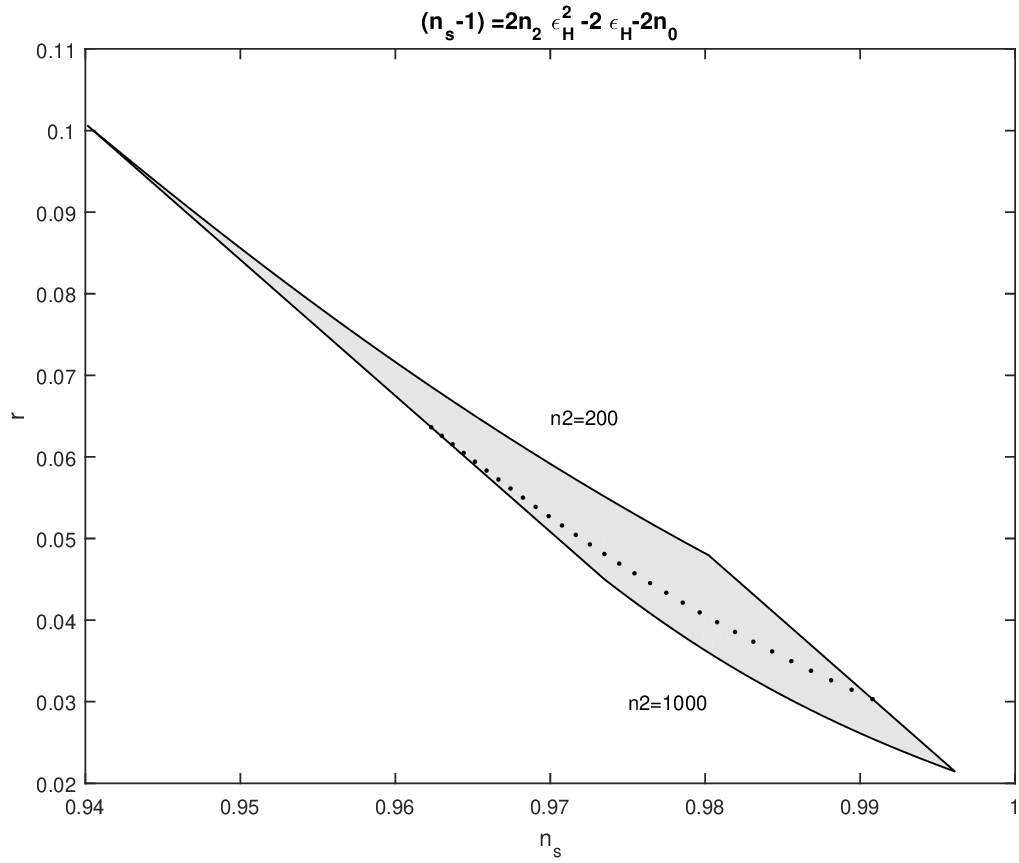}
%\mbox{\epsfxsize=14.2cm \epsffile{dustplusminus.eps}}
\caption{Spectral index $n_{s}$ to scalar to tensor ratio $r\,,$ for the
scalar field potential in which $n_{s}-1=2n_{2}\left(  \varepsilon_{H}\right)
^{2}-2\varepsilon_{H}-2n_{0}.~$The figure is for various values of \ the
parameter $n_{2}\in\left[  2\times10^{2},10^{3}\right]  $ and $n_{0}\in\left[
10^{-4},0.3\right]  $ while for the number of e-folds we sellected $N_{e}=55$.
The dot line is for $n_{2}=5\times10^{2}$. }%
\label{fig06}%
\end{figure}

Similarly, for the solution in which $n_{0}\neq0$ but $n_{1}=1$; that is, for
the expression (\ref{in.53}), we omit the derivation of the parameters
$n_{s}-r$. However, in Fig. \ref{fig06} the $n_{s}-r$ diagram is presented for
a number of e-folds given by $N_{e}=55$ and $n_{0}\in\left[  10^{-4}%
,0.3\right]  $ and $n_{2}=\left[  2\times10^{2},10^{3}\right]  $.

\section{Conditions to escape from Inflation}

\label{escape}

It is an open question as to which values for the free parameters of our
models determine when inflation ends. In order to answer this, we consider the
master equation (\ref{in.48}) and specifically we choose to rewrite it in
terms of the HSR parameter $\varepsilon_{H}\left(  \omega\right)  $ as,%
\begin{equation}
3\varepsilon_{H}^{\prime}=\left(  n_{0}+\left(  n_{1}-1\right)  \varepsilon
_{H}-n_{2}\varepsilon_{H}^{2}\right)  \varepsilon_{H}.\label{sc1}%
\end{equation}
This equation has the following critical points:
\begin{equation}
\varepsilon_{H}^{\left(  0\right)  }=0~,~\varepsilon_{H}^{\left(  \pm\right)
}=\frac{n_{1}-1\pm\sqrt{\left(  1-n_{1}\right)  ^{2}+4n_{0}n_{2}}}{2n_{2}%
}~,\text{for}~n_{2}\neq0,\label{sc2}%
\end{equation}
or%
\begin{equation}
\varepsilon_{H}^{\left(  0\right)  }=0~,~\varepsilon_{H}^{\left(  1\right)
}=\frac{n_{0}}{1-n_{1}}~,\text{when}~n_{2}=0~\text{and }n_{1}\neq1.\label{sc3}%
\end{equation}
Hence, in order for inflation to end in the cosmological models that we
studied, the free parameters of the models have to be constrained so that one
of the critical points, $\varepsilon_{H}^{\left(  \pm\right)  }$ or
$\varepsilon_{H}^{\left(  1\right)  }$, is an attractor, and also that
$\varepsilon_{H}^{\left(  \pm\right)  }\geq1$ or $\varepsilon_{H}^{\left(
1\right)  }\geq1$. We note that point $\varepsilon_{H}^{\left(  0\right)  }$
describes a de\ Sitter universe (that is, $w_{\phi}=-1$), while for the other
critical points the equation of state parameter, $w_{\phi}$, is constant.
Therefore, from the previous analysis we see that at the critical points the
scalar field potential is described by the exponential function.~

We proceed by considering the cases (a) $n_{2}=0$ and (b)~ $n_{2}\neq0$, where
the number of critical points differs.

\subsection{Subcase $n_{2}=0$}

Let as assume the simple case which corresponds to the master equation
(\ref{in.40}); that is, $n_{2}=0~$and we assume that $n_{1}\neq1$. In that
consideration, the critical points of the system are the $\varepsilon
_{H}^{\left(  0\right)  }~$and $\varepsilon_{H}^{\left(  1\right)  }~$of
(\ref{sc3}).

As far as concerns the stability of these points, we find that point
$\varepsilon_{H}^{\left(  1\right)  }$ is the unique attractor of the equation
when $n_{0}>0$, and $\varepsilon_{H}^{\left(  1\right)  }$ describes a point
without acceleration when $n_{1}<1$ and $n_{0}>1-n_{1}$. On the other hand,
when $n_{0}<0$, the unique attractor of the system is the de Sitter point
$\varepsilon_{H}^{\left(  0\right)  },$ although in this case the model does
not provide an exit from inflation.

\subsection{Subcase $n_{2}\neq0$}

For $n_{2}\neq0,$ a necessary condition for an exit from the inflation to
occur, is that the critical points $\varepsilon_{H}^{\left(  \pm\right)  }$
are real; that is, $4n_{0}n_{2}\geq-\frac{\left(  1-n_{1}\right)  ^{2}}{4}$.
\ In the special limit in which $n_{0}=0$, the points $\varepsilon
_{H}^{\left(  \pm\right)  }$ reduce to $\varepsilon_{H}^{\left(  0\right)  }$
and $\varepsilon_{H}^{\left(  2\right)  }=\frac{n_{1}-1}{n_{2}}$. \ In that
case, the two points are stable when $n_{2}>0$, and $\varepsilon_{H}^{\left(
2\right)  }$ is positive for any value of $n_{1}>1$.

In the general scenario with $n_{0}\neq0,$ it follows easily that in order for
$\varepsilon_{H}^{\left(  0\right)  }$ to be an elliptic point we require
$n_{0}>0$. Moreover, by assuming the condition $\varepsilon_{H}^{\left(
\pm\right)  }>1,$ we find that only the point $\varepsilon_{H}^{\left(
+\right)  }$ can be an attractor outside the inflationary era and this is
possible only when the free parameters satisfy the conditions
\begin{equation}
(i)~n_{2}<0,~n_{1}<1+2n_{2},~n_{0}>1-n_{1}+n_{2}\text{ \ and \ }4n_{0}%
n_{2}\geq-\frac{\left(  1-n_{1}\right)  ^{2}}{4},
\end{equation}
or
\begin{equation}
(ii)~n_{2}>0,~n_{1}>1+n_{2}~\text{\ and \ }n_{0}>1-n_{1}+n_{2},
\end{equation}
or
\begin{equation}
(iii)~n_{2}>0,n_{1}\leq1+n_{2}~\text{\ and \ }n_{0}>0\,.
\end{equation}
\ Hence, for values of the free parameters in those ranges only the third
model, i.e. where $h\left(  r\right)  $ is a quadratic function, admits an
attractor outside the inflationary era.

\begin{figure}[ptb]
\includegraphics[height=6cm]{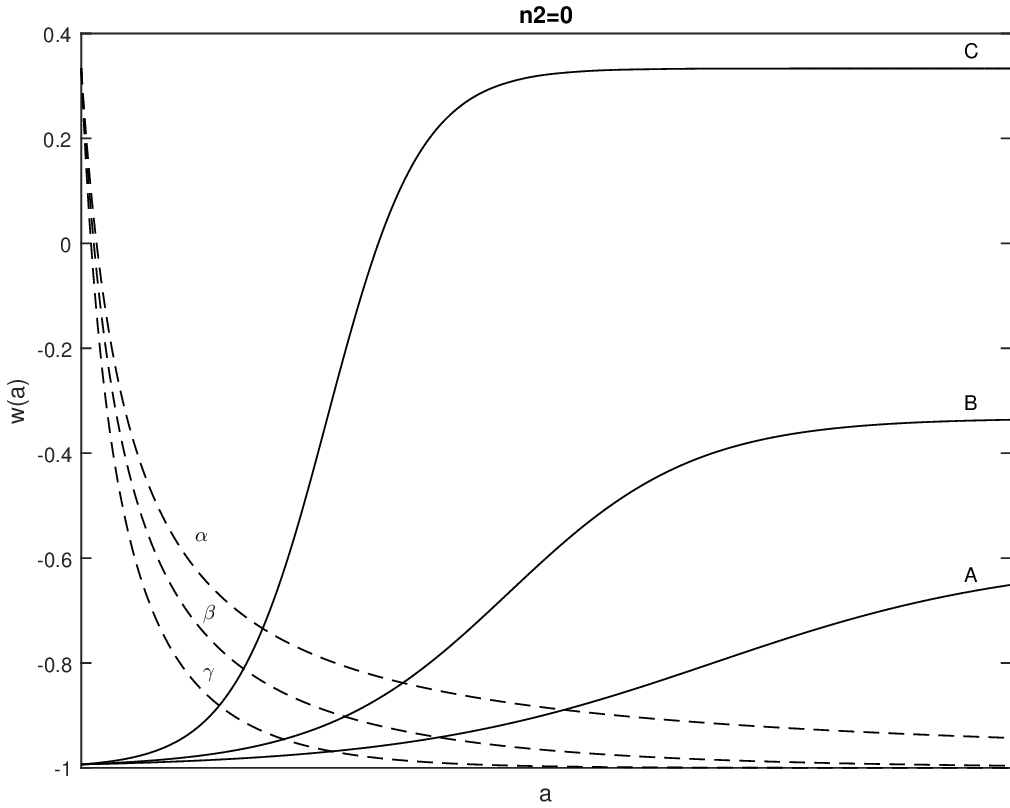}
\includegraphics[height=6cm]{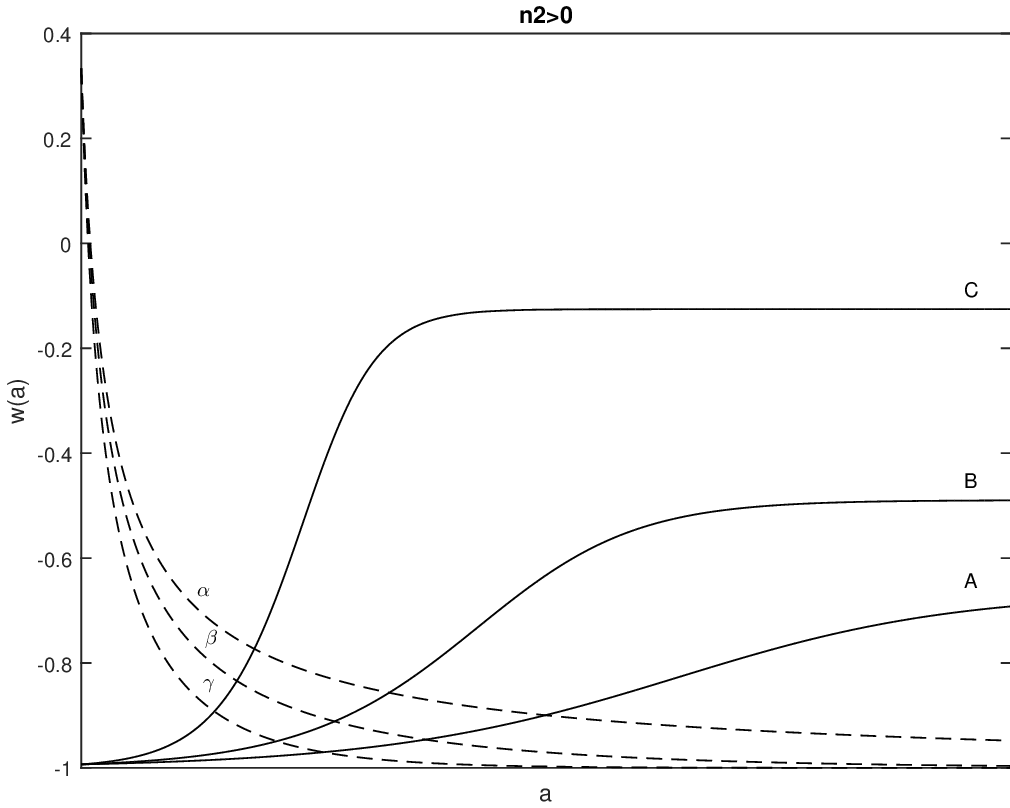}
\includegraphics[height=6cm]{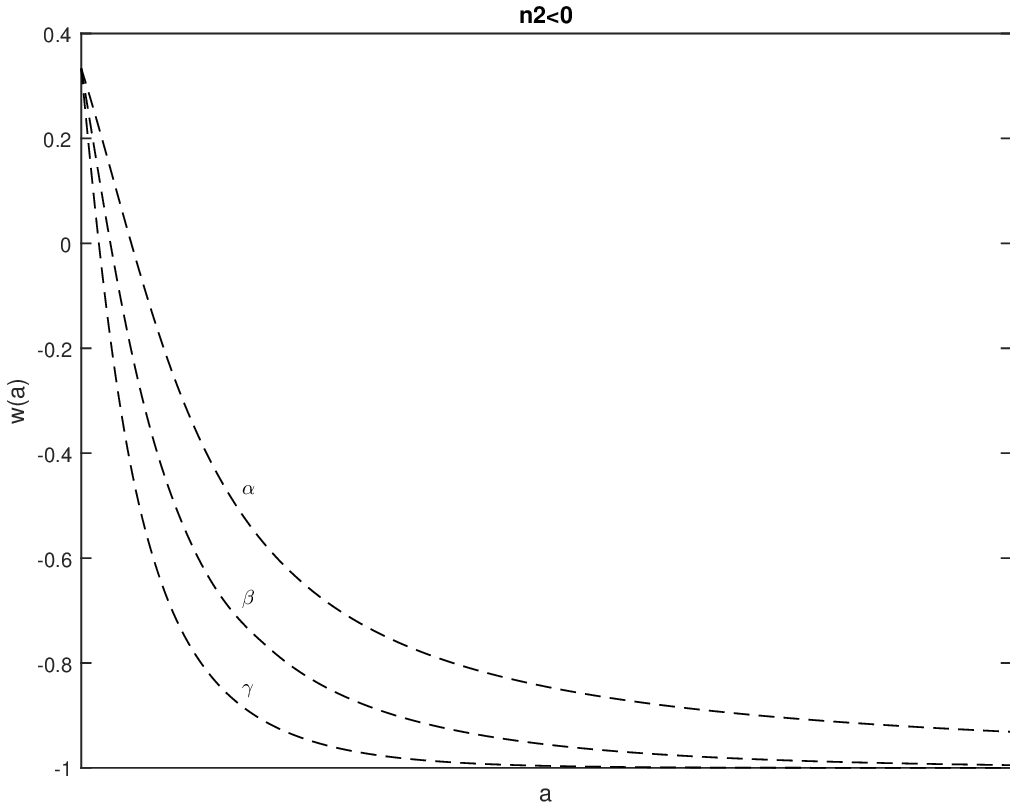}
\includegraphics[height=6cm]{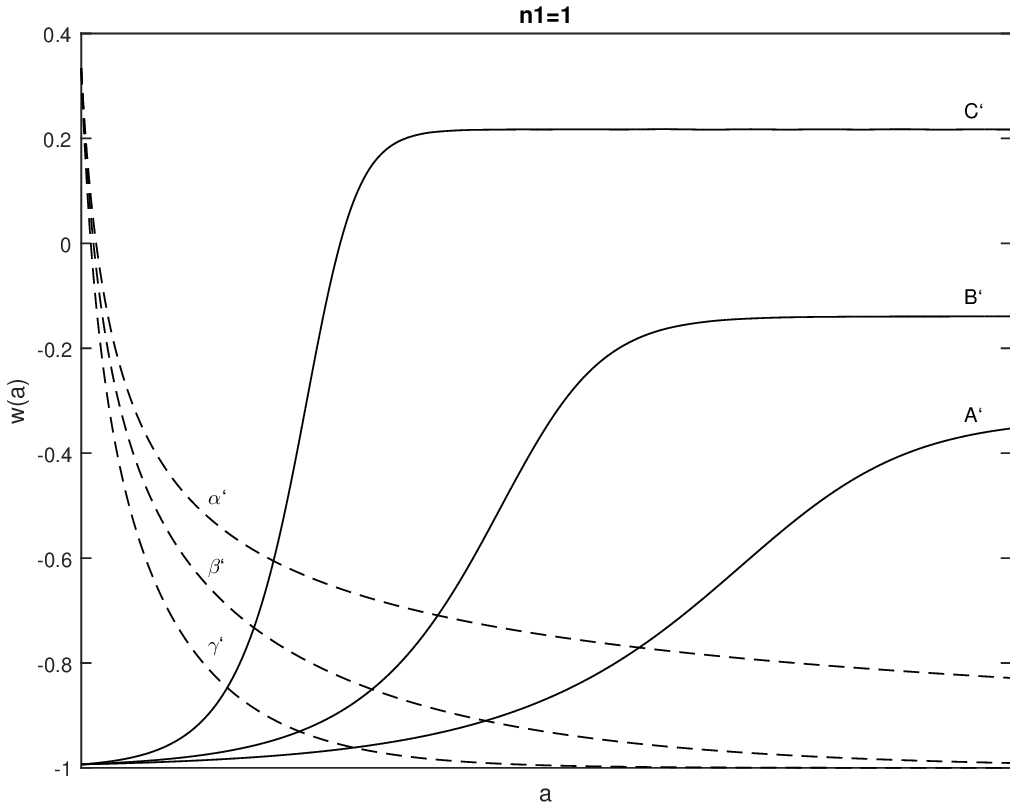}
%\mbox{\epsfxsize=14.2cm \epsffile{dustplusminus.eps}}
\caption{Qualitative evolution of the equation of state parameter $w\left(
a\right)  $, given by the solution of equation (\ref{sc1}). The solid lines
are for initial condition $\varepsilon_{H}\left(  a_{0}\right)  =0.01$ while
the dash-dash lines are for initial conditions $\varepsilon_{H}\left(
a_{0}\right)  =2$. As far as concerns the parameters $\mathbf{n}\left(
X\right)  =\left(  n_{0},n_{1}\right)  $ we have that $\mathbf{n}\left(
A\right)  =\left(  0.3,0.5\right)  ,~\mathbf{n}\left(  B\right)  =\left(
0.5,0.5\right)  $, $\mathbf{n}\left(  C\right)  =\left(  1,0.5\right)
,~\mathbf{n}\left(  \alpha\right)  =\left(  -0.1,0.5\right)  $, $\mathbf{n}%
\left(  \beta\right)  =\left(  -0.2,0.5\right)  $ and $\mathbf{n}\left(
c\right)  =\left(  -0.5,0.5\right)  $. Upper-left fig. is for $n_{2}=0$,
upper-right fig. is for $n_{2}=0.2$, and lower-left fig. is for $n_{2}%
=-0.2.~$Furthermore, the free parameters on the lower-right figure are
$\mathbf{n}^{\prime}\left(  X\right)  =\left(  n_{0},n_{2}\right)  ,$ such
that $\mathbf{n}\left(  A\right)  =\left(  0.3,0.3\right)  ,~\mathbf{n}\left(
B\right)  =\left(  0.5,0.3\right)  $, $\mathbf{n}\left(  C\right)  =\left(
1,0.3\right)  ,~\mathbf{n}\left(  \alpha\right)  =\left(  -0.1,0.3\right)  $,
$\mathbf{n}\left(  \beta\right)  =\left(  -0.2,0.3\right)  $ and
$\mathbf{n}\left(  c\right)  =\left(  -0.5,0.3\right)  $ while $n_{1}=1$. The
values of the free parameters have been chosen such that to cover the
stability analysis of equation (\ref{sc1}).}%
\label{figww}%
\end{figure}

In Fig. \ref{figww} the qualitative evolution of the equation of state
parameter $w\left(  a\right)  $, given by the solution of equation (\ref{sc1})
is presented for various values of the free parameters.

\section{Equivalent transformations}

\label{section4}

It is interesting that, when we set $n_{s}-1=0$, the scalar field mimics the
generalized Chaplygin gas (\ref{in.27}) with $\lambda=2$. Yet, when we assumed
that $\lambda\neq2$ in the equation of state of the generalized Chaplygin gas,
we found that $n_{s}-1=-2n_{1}\varepsilon_{H}$, where $\lambda=2-n_{1}$. These
two models are the solutions of the two different master equations,
(\ref{in.25}) and (\ref{in.35}), respectively. Yet, these two equations are
different for $n_{1}\neq0$, we observe that there exists a transformation
$F\left(  \omega\right)  \rightarrow\bar{F}\left(  \omega\right)  $, allowing
equation (\ref{in.25}) to be written in the form of (\ref{in.35}) and vice versa.

Suppose that $n_{1}\neq0,1$, then if in (\ref{in.25}) we substitute
\begin{equation}
F\left(  \omega\right)  \rightarrow\left(  1-n_{1}\right)  \bar{F}\left(
\omega\right)  , \label{in.59}%
\end{equation}
equation (\ref{in.25}) becomes
\begin{equation}
\bar{F}^{\prime\prime}+\left(  1-n_{1}\right)  \left(  \bar{F}^{\prime
}\right)  =0 \label{in.60}%
\end{equation}
which is just equation (\ref{in.30}). The transformation alters the line
element of the \ FLRW spacetime (\ref{in.14}) to
\begin{equation}
ds^{2}=-\left(  e^{-\bar{F}\left(  \omega\right)  }\right)  ^{\left(
1-n_{1}\right)  }d\omega^{2}+e^{\omega/3}(dx^{2}+dy^{2}+dz^{2}). \label{in.61}%
\end{equation}

A similar observation holds for the master equations (\ref{in.30}) and
(\ref{in.40}). Under the transformation, (\ref{in.59}), these two equations
are related, so a known solution for the model with EoS (\ref{in.42a}) for a
specific $\lambda$ can be used to construct a solution for another
cosmological model with a similar EoS parameter but with some other constant
$\lambda$. For completeness, note that in the case of $n_{1}=1$, the
transformation which relates the different set of equations is not that of
(\ref{in.59}) but $F\left(  \omega\right)  =\ln\left(  \bar{F}\left(
\omega\right)  \right)  $. \

On the other hand, it is important to mention that equation (\ref{in.35}) can
be written in the form of (\ref{in.25}) under the simple change of variable
$\omega=\frac{3}{n_{0}}\ln\left(  \bar{\omega}\right)  $. The same
transformation can be applied in the master equation, (\ref{in.40}), which is
transformed into equation (\ref{in.35}). Moreover, if we also apply the
transformation (\ref{in.59}) to (\ref{in.40}), then the latter takes the form
of the master equation, (\ref{in.25}).

These two transformations modify the line element of the FLRW
spacetime\ (\ref{in.14}) to%
\begin{equation}
d\bar{s}^{2}=-\frac{9}{\left(  n_{0}\right)  ^{2}\omega^{2}}\left(
e^{-\bar{F}\left(  \bar{\omega}\right)  }\right)  ^{\left(  1-n_{1}\right)
}d\bar{\omega}^{2}+\left(  \bar{\omega}\right)  ^{1/n_{0}}(dx^{2}%
+dy^{2}+dz^{2}). \label{in.62}%
\end{equation}

Moreover, in the limit for which $n_{1}=1$ in (\ref{in.35}) the latter becomes
the equation of a free particle, while the resulting scalar field theory is
that of the exponential potential where the scalar field has a constant EoS parameter.

The existence of transformations of this kind, which transform the one model
into another, is not a coincidence. The master equations (\ref{in.25}),
(\ref{in.30}), (\ref{in.35}) and (\ref{in.40}) are maximally symmetric. In
particular they are invariant under the action of one-parameter point
transformations (Lie point symmetries) which form the $SL\left(  3,R\right)  $
Lie algebra. \footnote{According to Lie's Theorem,\ any second-order equation
which admits the elements of the $SL\left(  3,R\right)  $ algebra as
symmetries is equivalent to the equation of a free particle and all the
maximally symmetric equations commute \cite{sLie}. The map is the one which
transforms the admitted $SL\left(  3,R\right)  $ Lie algebra among the
different representations of the admitted equations, for more details see
\cite{lie1}.}

Consider now the classical Newtonian analogue of a free particle and an
observer whose measuring instruments for time and distance are not linear. By
using the measured data of the observer we reach in the conclusion that it is
not a free particle. On the other hand, in the classical system of the
harmonic oscillator an observer with nonlinear measuring instruments can
conclude that the system observed is that of a free particle, or that of the
damped oscillator or another system. From the different observations, various
models can be constructed. However, all these different models describe the
same classical system and the master equations are invariant under the same
group of point transformations but in different parametrization.

In the master equations that we studied there is neither position nor time
variables: the independent variable is the scale factor $\omega=6\ln a$, and
the Hubble function is the dependent variable, $H\left(  a\right)  $.
Therefore, we can say that at the level of the first-order approximation for
the spectral indices, various representations of the variables $\left\{
a,H\left(  a\right)  \right\}  $ provide different observable values for the
spectral indices. This property is violated when we consider the second-order approximation.

Transformations of this kind are well-known in physics. For instance, the
Darboux transformation for the Schr\"{o}dinger equation \cite{darboux} is just
a point transformation that relates linear equations with maximal symmetry;
that is,\ it belongs to exactly the same category of transformations that we
discuss here. A special characteristic of the Darboux transformation is that
it preserves the form of the equation but the potential in the Schr\"{o}dinger
equation changes. An application of the Darboux transformation for the
determination of exactly solvable cosmological models can be found in
\cite{darboux22}.

Transformations which keep the form of our master equation exist. We do not
have potential terms in the master equations but there are transformations
which change the constant coefficients appearing there while retaining the
form of the master equations.

In order to demonstrate this, consider the master equation (\ref{in.40}). The
application of the first transformation $F\left(  \omega\right)
\rightarrow\frac{1-\bar{n}_{1}}{1-n_{1}}\bar{F}\left(  \omega\right)  ~$in
(\ref{in.40}), preserves the form of the master equation but the constant
$\lambda$ in the equation of state for the generalized Chaplygin gas
(\ref{in.42a}) shifts from $\lambda=2-n_{1}$ to $\bar{\lambda}=2-\bar{n}_{1}$.
\ Moreover, the application of the second transformation, $\omega
\rightarrow\frac{\bar{n}_{0}}{n_{0}}\bar{\omega},$ in (\ref{in.40}) gives%

\begin{equation}
\frac{d^{2}\bar{F}}{d\bar{\omega}^{2}}+\left(  1-\bar{n}_{1}\right)  \left(
\frac{d\bar{F}}{d\bar{\omega}}\right)  ^{2}-\frac{\bar{n}_{0}}{3}\frac
{d\bar{F}}{d\bar{\omega}}=0 \label{in.63}%
\end{equation}
which is exactly the same master equation, just with different coefficients.

Furthermore, for the more general case that we studied (the master equation of
eqn.(\ref{in.48})) it is easy to see that for $n_{2}n_{0}\neq0$, eqn.
(\ref{in.48}) admits eight Lie point symmetries; that is, it is maximally
symmetric. Hence, there exists a mapping $\left\{  \omega,F\left(
\omega\right)  \right\}  \rightarrow\left\{  \Omega,\Phi\left(  \Omega\right)
\right\}  $ which transforms the master equation (\ref{in.48}) to that of a
free particle, or to any other maximally symmetric equation -- such as the
other master equations we studied above. \ Of course, this result can be used
to derive closed-form solutions in other models with a maximally symmetric
master equation.

Recall that a map in the space of the variables which transforms one solution
to any other solution was also found in \cite{new}. However, while both maps
transform solutions into solutions, the one that we have discussed here,
transforms not only solutions into solutions but systems of dynamical
equations into equivalent systems\footnote{Other transformations which belong
to these families of transformations are presented in \cite{charters}.}. In
order to reflect that latter property, the map is called an equivalent point transformation.

The elements of the $SL\left(  3,R\right)  $ -- except for the transformations
which relate algebraic equivalent equations -- provide us with important
physical information about the system under study. One of these properties
which arises from equation (\ref{in.25}) is the well-known scale invariance of
the Harrison-Zeldovich spectrum, regarding which it can easily be seen that
equation (\ref{in.25}) is invariant under transformations $\omega
=\omega^{\prime}+\omega_{0}$ or $\omega=\omega^{\prime}e^{\bar{\omega}_{0}}$,
where these two transformations are related with the symmetry vectors
$\partial_{\omega}$ and $\omega\partial_{\omega}$. In particular, every
element of the $SL\left(  3,R\right)  $ is related to a point transformation
which leaves the differential equation, and consequently the solution,
invariant. Moreover, with a different reparameterization of the $SL\left(
3,R\right)  $, for equivalent models, the physical interpretation of the
invariant point transformations can change between the different models.

\section{Conclusions}

\label{conc}

In scalar-field cosmology, the dark-energy EoS and the inflationary
scalar-field potential have been reconstructed from the spectral index,
$n_{s}$. From the Planck 2015 data analysis, it is known that the observable
variables -- the tensor-to-scalar ratio, $r$, and the spectral index for the
density perturbations, $n_{s}$ -- form a surface in the $n_{s}-r$ plane.
Furthermore, these two observable variables can be expressed in terms of the
slow-roll parameters and their derivatives. Therefore, the ansatz that the
spectral index for the density perturbations is related with the
tensor-to-scalar ratio, $\left(  n_{s}-1\right)  =h\left(  r\right)  $,
provides a differential (master) equation whose solution defines the
corresponding cosmological model.

In this paper, we assumed $n_{s}$ to be given in the first approximation by a
function $h\left(  r\right)  $ that it is: (a) constant, (b) linear, and (c)
quadratic, respectively. In order for the first-order approximation to be
valid the free parameters which have been introduced by the function $h\left(
r\right)  $ have to satisfy some consistency conditions.

We work with the HSR parameters. The case in which $h\left(  r\right)  $ is
constant, that is, $n_{s}-1=-2n_{0},\,$\ is one that has been studied before
in the literature and, in the limit, $n_{0}=0$, corresponds to the
Harrison--Zeldovich spectrum. The differential equation which follows provides
the scalar factor to be that of a specific intermediate inflation, $a\left(
t\right)  \simeq\exp\left(  a_{1}t^{2/3}\right)  $, while the corresponding
perfect fluid satisfies the equation of state (\ref{in.27}). On the other
hand, for nonzero $n_{0}$, we found that the scalar field satisfies an EoS
given by expression (\ref{in.42a}) for $\lambda=2,$ which includes expression
(\ref{in.27}). For the scalar-field potential, the construction looks similar,
and for $n_{0}=0$ the potential is given in terms of polynomials of the field
$\phi$, and for $n_{0}\neq0$ in terms of hyperbolic trigonometric functions.

As a second generalization, we assumed $h\left(  r\right)  $ to be the linear
function, $h\left(  r\right)  =-\frac{n_{1}}{5}r-2n_{0}$. Now, the models
derived from the differential equation $n-1=h\left(  r\right)  ,$ in the
first-order approximation, are the generalized Chaplygin gases, (\ref{in.27})
and (\ref{in.42a}), for $n_{0}=0$ and $n_{0}\neq0,$ respectively; where now
the power $\lambda$ in the equations of state is related to the value of
$n_{1}$, by $\lambda=2-n_{1}$.

Finally, the case in which $h\left(  r\right)  $ is a quadratic polynomial was
considered and two new equations of state which generalize the Chaplygin gas
were derived. Exact examples displaying a generalised sudden singularity of
the type identified by Barrow and Graham \cite{graham} for inflationary scalar
fields with fractional potentials were found here. Lastly, the ranges for the
values of the free parameters of the models have been considered which permit
the universe to escape from the inflationary phase.

It is important to mention that in this work we have assumed that we are in
the inflationary epoch and so the equation of state parameters, or
equivalently the scalar field potentials that we reconstructed, can be seen as
the leading order terms, or attractors, of a more general equation of state
parameter which describes the whole evolution of the universe.

It is particularly interesting that the master equations we derived in our
study are second-order differential equations of maximal symmetry. Hence, they
are invariant under the action of point transformations with generators given
by the elements of the $SL\left(  3,R\right)  $ algebra. Every master equation
defines a representation of the $SL\left(  3,R\right)  $ algebra and the map
which changes the representation transforms the master equation to the
corresponding master equation of another model. This relates explicitly the
form of the line elements for the various cosmological models. The
transformation which performs the change is a projective transformation in the
jet-space of the master equation; that is, a map in the space of the dependent
variable $F\left(  \omega\right)  $ and the spacetime variable $\omega$ -- we
recall that $dt=e^{-F\left(  \omega\right)  /2}d\omega$ and $a\left(
t\right)  =e^{\omega/6}$.

In a forthcoming work we will investigate whether the latter result can be
extended to the case in which the master equation,~$n_{s}-1=h\left(  r\right)
$, is defined by higher-order approximations for the spectral indices.

\begin{acknowledgments}
JDB is supported by the Science and Technology Facilities Council of the
United Kingdom (STFC). AP acknowledges the financial support of FONDECYT grant
no. 3160121. AP thanks the Durban University of Technology for the hospitality
provided while part of this work was performed.
\end{acknowledgments}

\end{document}